\documentstyle[12pt]{article}
\begin{document}
\centerline{ }
\vspace{7cm}

\centerline{\bf {{NONPERTURBATIVE METHODS IN KAON PHYSICS}}}
\centerline{\bf {{ (ASIDE FROM
THE LATTICE)}}}

\vspace{.7cm}

\centerline{\bf John F. Donoghue}
\centerline{Department of Physics and Astronomy}
\centerline{University of Massachusetts, Amherst MA 01003 USA}

\vspace{5cm}

 I discuss the progress in the use of analytic techniques for
low energy QCD, in particular as applied to kaon physics. These
methods are becoming increasingly powerful and we have gained a good
deal of control over the difficult hadronic interactions. There are
continuing developments, and I speculate on the ways that these
techniques may become yet more sophisticated in the future.\\
\vspace{1cm}
\vfill
Invited talk presented at the Workshop on Kaon Physics, Orsay, June,
1996, to be published in the proceedings.
\pagebreak

The title of this talk (suggested by the organizers) and the
scheduling of it immediately after the fine lattice review by Greg
Kilcup$^1$ suggest a competition between analytic and computer
methodologies, man vs. machine. This dichotomy has always reminded me
of the American folk tale of John Henry. The story is from a time
before the LEP tunneling machines, when tunnels were dug by men
swinging 20 pound hammers. John Henry was the best of these until one
day someone arrived with a machine to do the job, and challenged
John Henry to a contest. John Henry replied that a man is only a man,
but that he would accept the challenge. The event unfolds with the
machine pulling ahead, whereupon John Henry grabs a second hammer and,
swinging one in each hand, begins to catch up. The machine eventually
breaks down, John Henry keeps going and he wins at the end of the day.
Unfortunately, the effort was too much even for the big heart of John
Henry, and he collapses and dies on the spot. At this stage, all
versions have the refrain that ``he died with a hammer in his hand''.
As a child, I used to think that the dying spoiled the story. However
as adults, knowing what we do about men and machines, it is clear that
dying is the only possible ending. The nobility was not the victory
but the effort - that he died with a hammer in his hand. For perhaps
obvious reasons, this story resonates with those of us who work with
analytic techniques. In any case, it is my task to take the side of
``man'', and I am pleased to report that it is still a contest. 

Our goal is to tame the low energy strong interactions in order to be
able to make predictions for the weak decays of kaons, with the hope
of extracting information about CP violation or rare weak interaction
processes. However, at low energies one cannot use QCD perturbation
theory so that these processes have always posed challenges. We do 
have at our disposal a set of rigorous techniques, and these have
allowed us to make some real progress in the treatment of low energy
physics. 
The analytic methods that are presently employed have an extended
history going back to the sixties. Although in their modern
incarnations they are more powerful than previously, I will review
briefly their roots, using a few papers from 1967 as examples$^2$. Then I
turn to our modern usage, and to our hope for the future. I would like
to convince you that, when looked at with the right tools, this
physics is not too complicated, and that we have hopes to produce more
solid calculations in the future.

\section{Roots}
a) Chiral Symmetry:

The first of our ``root'' examples is a seminal paper by
C. Bouchiat and Ph. Meyer$^3$ relating the reactions
$K \to 3 \pi$ to $K \to 2\pi$. They predict that the $\Delta I = 1/2$
amplitude for $K \to \pi^+ \pi^- \pi^0$ is 
\begin{equation}
A(K^0_L \to \pi^+\pi^-\pi^0) = {A_0 \over 6 F_\pi}\left[ 1 + 3 {(s_3 -
s_0)
\over m_K^2} \right]  \ , 
\end{equation}
\noindent where $s_3 = (k-p_0)^2$, $s_0 = m_K^2 + m_\pi^2/3$ and $A_0$
is the amplitude for $K_S \to 2 \pi$. The input to this result is
chiral symmetry, more specifically chiral SU(2) symmetry
which is an exact symmetry in the limit of vanishing up and down quark
masses$^4$.  
This may seem to be an unusual
type of prediction for a {\it symmetry}, in that it gives dynamical
information about the kinematic structure of the amplitude. 
In understanding why this occurs, one gets to the heart of chiral
techniques. Symmetries relate different states, with the most
familiar situation being the relation between single particle states.
For example, isospin symmetry relates the properties of neutrons and
protons, which live in the same isospin multiplet. However for the
axial vector rotations inherent in chiral symmetries, there are no
multiplets of particles that are related by a symmetry transformation.
Instead, if you transform a proton state you obtain a state
consisting of a proton plus a zero-energy pion. Therefore the typical
predictions of axial rotations are those relating a process to one
with an extra pion. If one provides a Taylor expansion of the
amplitude in terms of the various particle momenta, the symmetry fixes
the zero-momentum limit. The specific technique used in this context
are the so-called soft-pion theorems, where ``soft'' refers to the
zero-momentum limit. In the case of $K \to 3\pi$ there are three
separate zero momentum limits. The Bouchiat Meyer calculation allows
an expansion to two powers of the external momenta, in which case the
three limits are sufficient to completely determine the amplitude. 

This allows us also to understand the nature of the corrections to the
calculation. A simple class of corrections
 would occur because the chiral SU(2)
symmetry is broken by the u,d quark masses, which would lead to
corrections to the prediction by higher powers of $m_\pi^2$. 
However this is not the dominant effect. There are also corrections
which come from the fact that, in the Taylor expansion of the
amplitude, we stopped at a quadratic momentum dependence. There are
quartic  terms, such as $k \cdot p_0 p_+ \cdot p_- $, which vanish in
all zero-momentum limits  and hence have a coefficient which is not
fixed by the symmetry alone. The kinematic factor cited in the
previous sentence  is of order $m_K^4$ compared to terms of order
$m_K^2$ from the quadratic terms so that the correction is of relative
order $m_K^2/\Lambda^2$, where $\Lambda$ is some scale. In practice,
$\Lambda$ is of order 1 GeV. 

This pattern is typical of chiral relations. One gets real dynamical
information about processes with differing numbers of pions. The
lowest order momentum dependence is often fixed by the symmetry in
terms of another reaction. Then there occur unknown coefficients
corresponding to higher order momentum dependence. Sometimes one can
in turn relate these coefficients to ones measured in other
processes.  In the context of this talk it is useful to point out that
these results are fully nonperturbative, and constitute low energy
theorems of QCD. \\

\noindent 2) Effective Lagrangians 

A 1967 paper by Jeremiah Cronin$^5$ provides a good example of an
effective way to organize the predictions of chiral symmetry. By
writing a Lagrangian involving the pion fields which has 
the chiral symmetry, one can easily read off the symmetry predictions
in a way much simpler than using the soft pion theorem. This can work
because if the predictions are to be consequences of the symmetry
alone, then all Lagrangians with the same symmetries will share the
same predictions. Since the chiral predictions involve differing
numbers of pions, and this process can be continued until any number
of pions are present, the effective Lagrangian for chiral chiral
symmetry must be nonlinear, involving all numbers of pion fields.
Cronin's paper contains the Lagrangian for kaon decays which yields
the result quoted above

\begin{equation}
L_W = g_8 Tr \left( \lambda_6 D_\mu U D^\mu U^\dagger \right)
\end{equation}
\noindent where
\begin{equation}
U = exp \left( i {\lambda^A \phi^A \over F_\pi}\right)  .
\end{equation}
[Technically, the soft pion theorems use chiral SU(2) while this
Lagrangian is defined in chiral SU(3). In this case, the
difference is not particularly important.]
This Lagrangian is determined by the symmetry of the weak
interactions, in which only left-handed fields participate in the
charged current processes, plus the fact that we know that the  
octet nonleptonic interaction is much larger than the 27-plet. [A
similar Lagrangian can be written for the 27-plet.]

One interesting extension of the use of effective lagrangians is to
include the consequences of the corrections to the lowest order chiral
relations, such as I described above as coming from quartic terms in
the Taylor expansion of the decay amplitude. There exist effective
Lagrangians which have the same symmetry as the one in Eq.3, but which
have more derivatives. An example is 

\begin{equation}
L_{h.o.} = g' Tr \left( \lambda_6 D_\mu U D^\mu U^\dagger D_\nu U 
D^\nu U^\dagger \right)
\end{equation}
Because it has four derivatives on the meson fields, it will lead to
corrections which have four powers of the momenta, 
such as were described above. The coefficient of such a
Lagrangian would not be known ahead of time, but similar four
derivative Lagrangians could be useful in categorizing the deviations
form the lowest order predictions. Chiral symmetry constrains not only
the zero-momentum limit, but also the corrections to that limit. This
latter role is much easier to study systematically using effective
Lagrangians. \\

\noindent 3) Dispersion relations 

Another topic which was active both in 1967 and at present is the use
of dispersion relations as a calculational tool. Examples of this are
the Weinberg sum rules$^6$ and the calculation of the pion
electromagnetic mass difference by Das et al$^7$. In 1967 these
involved rather bold assumptions about the short distance/high energy
behavior of the theory, but we now know that these assumptions are
satisfied in QCD with massless quarks. These calculations involve the
vector and axial vector spectral functions 
$\rho_V$ and $\rho_A$, which are measurable in
$e^+ e^- $ annihilation and in $\tau$ decays, and whose high energy
behavior is known in QCD. Specifically, we have 
the Weinberg sum rules
\begin{equation}
F^2_{\pi} = \int^{\infty}_0 ds ( \rho_V (s) - \rho_A (s))
\end{equation}

\begin{equation}
0 = \int^{\infty}_0 ds s ( \rho_V (s) - \rho_A (s))
\end{equation}
\noindent as well as a sum rule for one of the chiral parameters
\begin{equation}
-4 \bar{L}_{10} = \int^{\infty}_{4m^2_{\pi}} {ds \over s} ( \rho_V (s) - 
\rho_A (s))
\end{equation}

\noindent with

\begin{eqnarray}
\bar{L}_{10} &=& L^r_{10} (\mu) + {1 \over 192 \pi^2} \left[ ln 
{m^2_{\pi} \over \mu^2} + 1 \right] \\ \nonumber
&=& (-0.7 \pm 0.03) \times 10^{-2} \; (Expt: \pi \rightarrow e \nu \gamma)
\end{eqnarray}
\noindent and a sum rule for the pion electromagnetic mass difference
\begin{equation}
m^2_{\pi^+} - m^2_{\pi^0} = -{3e^2 \over 16 \pi^2 F^2_{\pi}} 
\int^{\infty}_{0} ds s ln s \left[ \rho_V (s) - \rho_A (s) \right].
\end{equation}

\noindent These are theorems of QCD in the chiral limit. The first and
third of these
remain true even in the presence of quark masses, but the other two
are no longer convergent. The spectral functions are pretty well
known$^8$ and are consistent with these sum rules. 

Dispersion relations derive their validity from the general analytic
properties of amplitudes in field theory. They relate the real and
imaginary parts of various amplitudes, with the general structure
\begin{equation}
f(s) = {1 \over \pi} \int_{0}^{\infty} ds' {Im f(s') \over s' - s - i 
\epsilon}
\end{equation}
\noindent or possibly with subtractions to make them convergent. The imaginary
parts correspond to real on-shell intermediate states, and can often
be measured directly by experiment. These can then be used to predict
the full amplitude at all values of the kinematic variables.

The importance of this technique lies in the fact that dispersion
relations are the only rigorous analytic methods that is 
able to handle the intermediate energy
regions in QCD. However, of course, the output of a dispersion
relation is only as good as the input, and only sometimes we do know the
relevant imaginary part of the amplitude.

\section{Effective field theory}
The techniques described above form the basis for much of the work on
low energy QCD which is required for the study of kaon interactions. 
However, there is one very important extra ingredient which was not
present in 1967 but which is a key to modern applications. This
technique is effective field theory$^9$. 

Most of us learned field theory in the context of renormalizable
field theories, and these are certainly attractive for the fundamental
interactions. However at low energies one is often faced with an
effective theory which need not be renormalizable. Do we have to solve
the full high energy theory in order to know about quantum predictions
at low energy? Intuitively we know that the most important physics at
low energies is that involving whatever light degrees of freedom are
present in the low energy theory. This is why condensed matter physics
or atomic physics can proceed without knowing about quarks and gluons.
Effective field theory is a formalism which has been developed to
handle such a situation. It allows one to calculate the quantum effects
of the light degrees of freedom and encodes one's ignorance of the
ultimate high energy theory into a set of parameters in an effective
Lagrangian. It then brings the full field theoretic apparatus to bear
on non-renormalizable theories when treated at low enough energies.

The importance of this for low energy QCD is that it converts the
constraints of chiral symmetry into a fully dynamical field theory,
called chiral perturbation theory$^{4,10}$, which 
one can justifiably claim is
rigorously equivalent to QCD when applied at low energy. Chiral
symmetry dictates the low energy couplings of the light pions and
kaons, and the propagation and rescattering of these are calculated in
effective field theory. The usual infinities of perturbation theory
are absorbed into the unknown coefficients of the effective
Lagrangian, and the renormalized values of these must be determined
from experiment. This gives a systematic method for calculating not
only the lowest order amplitude but higher order corrections in the
energy expansion as well as quantum corrections. Chiral perturbation
theory is now a well developed technology, and serves as a model for a
complete effective field theory in much the way that QED serves as a
model for a complete renormalizable field theory.

\section{Where are we now?}
The development of chiral perturbation theory has radically
transformed our treatment of low energy QCD. It has given us a
calculational tool which has rigor and which is useful
for phenomenology. Although there are some important limitations to 
the utility of the method, it has justifiably become the standard
basis for calculations of low energy reactions.
 
 The present frontier in mesonic chiral perturbation theory
involves two-loop calculations. These now exist in the literature
 for several
quantities$^{11,12,13}$. A two-loop 
calculation is equivalent to an order $E^6$ in
the energy expansion, and brings in new chiral parameters from the
Lagrangian at that order. The special physics that is revealed in
these calculations appears to me to be the iteration of final state
rescattering. In several reactions, especially involving the $I=0$,
$J=0$ two pion state, the rescattering is quite strong at one-loop and
thus it is relevant to calculate the two-loop effect. My favorite
example is the reaction $\gamma \gamma \to \pi^0 \pi^0$ studied by
Bellucci, Gasser and Sanio$^{11}$. There is no tree level contribution at
orders $E^2$ and $E^4$, although there is a finite one loop effect
which reflects a difference in the $\pi^+ \pi^- \to \pi^0\pi^0$
rescattering in the $I=0$ and $I=2$ channels. The one loop result is
a bit low even near threshold, but the two-loop result nicely corrects
this flaw. The important ingredient near threshold does appear to be
the more correct treatment of the large $I=0$ rescattering, as the
threshold region is quite insensitive to any new chiral parameters. 
In general, however, most two-loop calculations will always have some model
dependence, at least at higher energies, in that the new chiral
parameters will not be able to determined experimentally for use in
making predictions (there are too many of them), so that they will
need to be estimated using less rigorous models. 

Another development in recent years is the increased use of dispersive
techniques in connection with chiral calculations. I have reviewed
this in greater depth elsewhere$^{14}$, so that I will mainly state the main
points here. Any chiral loop calculation can be reformulated as a
dispersion relation, since the Feynman diagrams have the same
analyticity structure assumed for the dispersion relation. Use of the
lowest order chiral prediction for the imaginary parts of diagrams 
reproduces the usual one-loop result. Then any improvement in the imaginary 
part, especially through the use of data, will lead to improved
predictions. The matching of the low and moderate energy regions is
known, with chiral symmetry providing information on the subtraction
constants and dispersion relations providing the extrapolation to
higher energy. Again  $\gamma \gamma \to \pi^0 \pi^0$ can provide a
nice example$^{15}$. By matching to the chiral calculation we remove all free
parameters in the dispersive treatment, and the dispersion
corrections provide the iteration of the rescattering diagrams to all
orders. Again some modest model dependence enters at high energies.
However, the  dispersive and two-loop calculations agree extremely
well, and I feel the the physics of this process is well under
control. The dispersive techniques involving a matching with chiral
perturbation theory are quite promising as calculational tools.
 
This technique also opens up completely new forms of calculations.
Here the Das et al$^{7}$ calculation of the pion electromagnetic mass
difference in the chiral limit from 1967 is a model. The
electromagnetic effect involves the integration over all loop momenta,
but the dispersive method converts this into a sum rule involving
$\rho_V(s)-\rho_A(s)$. Since we know these spectral functions well
enough, we can convert this into a calculation of the mass difference.
This type of calculation goes beyond what can be done in pure chiral
perturbation theory, where the electromagnetic mass difference is
described by a unknown parameter. This has also been used to calculate
a weak nonleptonic matrix element in the chiral limit$^{17}$.

\section{Man vs machine} 
    Computer methods are also used to address some of the same
problems we have described above. However, there are some constraints
on these methods, both temporary and long term. In the short term, the
quenched truncation is an issue of unknown severity, as it is not even
a valid approximation scheme in QCD (in the sense that there is no
small parameter such as $1/N_c$ that controls the size of
corrections). We know from chiral calculations that loop effects are
important in some matrix elements such as the B parameter-these are
misrepresented by the quenched truncation. The inability to reach very
low energies/masses is another short term problem. There is also a
more difficult issue that will always remain in that lattice results
are in the Euclidian region. For processes where physical intermediate
states play an important role, these effects will be missed by a
Euclidian simulation. It is not clear that this part of the
continuation to the Minkowski region can ever be built in in a
rigorous way. 

For these areas, the rigorous analytic methods described above are 
still superior to computer simulations. In fact, even in some
intermediate energy applications man may still do better than
machine. For example, in the dispersive calculation of the
weak matrix element mentioned above, the Monte Carlo methods can be
thought of as producing simulations to the relevant intermediate
states, while the dispersive method uses real data for the same. A
comparison here may be more a test of the Monte Carlo method.  

\section{Models}

Phenomenologically, the physics on the low energy region is not very
complicated. The structure of almost any amplitude as a function of
energy involves a few visible resonances merging into a high energy
continuum. Particularly clear examples of this are the vector and
axial-vector spectral functions, and the structure functions of deep
inelastic scattering. The primary resonances that are involved are few
in number, with the rho playing the prominent role and the resonances
up to 1.4 GeV occasionally being visible. The simplicity of this physics has
led to the development of models which attempt to provide a
useful description of this dynamics. These models invariably drop
some aspects of the full dynamics, and so they are not rigorous
techniques. However, to the extent that they capture some of the
correct physics, they may be convenient and reasonably accurate ways
to handle the intermediate energy region.

\noindent a) Vector Dominance: 

Vector dominance or resonance saturation may be considered as a poor
man's dispersion relations. If one takes a dispersion relation and
replaces the integrand by a zero-width resonance, one obtains the
resonance saturation approximation. As noted above, this most often
involves vector mesons such as the rho. The result can be turned onto
a field theory calculation, with the various transitions being
described by measured coupling constants in a Lagrangian. The specific
application of vector dominance turns out to be surprisingly subtle,
keeping referees and authors busy correcting the multiple mistakes
which are possible. However, done properly it does capture a good deal of the
right low energy physics. For example, the major chiral parameters in
the chiral effective Lagrangian can be predicted by resonance
saturation$^{18}$.

\noindent b) Quark Loops:
 
Resonances are reasonably well understood as $q\bar{q}$ bound states,
and so intermediate states with a resonance could be thought of as
having  $q\bar{q}$ propagation. An extreme limit of this propagation
is a free quark loop$^{19}$, and this model has been used to compute various
low energy processes. When thought of in the resonance plus continuum
language this limit is the reverse of vector dominance, being all
continuum and no resonance. In practice, this model is less successful
than vector dominance, but still it does surprisingly well considering
the naivity of the approximation.

\noindent c) NJL models:

There is a whole subfield devoted to Nambu Jona-Lasinio 
models$^{20,21}$. The
idea is to include a four-quark contact interaction to model the
QCD interaction between quarks. Using this with a cutoff produces a
model with a light pion with nontrivial chiral interactions. Of all
the variants of this idea, my favorite is that of Ref 21, because they
use it in a way that connects easily to both chiral and dispersion
techniques. By summing up classes of diagrams, they generate
intermediate states that contain both resonance-like bumps and
continuum contributions (see for example the vector spectral function
of Ref 21. ). This is approaching the right physics. The cutoff may
sometimes make it difficult to perform a valid matching to high
energy, but this is certainly an improvement over the use of free
quark loops.  
    
None of these models is yet the full story. We need a way to capture
both resonance and continuum physics as accurately as possible. My
feeling is that this is best done in a dispersion theory context. 

\section{Where are we going?}
There are clearly many applications of analytic methods to specific
reactions, and more will be studied in the future. However, my focus
here is not on these but on the way that techniques as a whole are
developing. To my biased eye, there actually is new direction in
techniques which has the potential for great importance if we can
develop it sufficiently. This involves the calculation of
``nonleptonic'' processes which involve current matrix elements
integrated over all scales. Such calculations are still in the
exploratory stages, but the issues that are being studied, such as the
matching of short and long distance physics, are the final frontier of
analytic methods.

Many of processes that we have presently mastered are matrix elements
of a single current. However, typical of the new more difficult class
is the electromagnetic mass differences of the mesons, which involves
two electromagnetic currents in which the intermediate states are
integrated over together with the photon propagator. Since these
intermediate states involve all scales, it is not sufficient to know
the short distance physics from  QCD perturbation theory, nor the long
distance physics from chiral perturbation theory. In addition to these,
we must be able to bring them together in the intermediate energy
region in an accurate way. This is the challenge.

The weak nonleptonic matrix elements are the most difficult of these.
The $\Delta I = 1/2 $ rule still has not been definitively explained in
a way that is convincing to the full community.
In some ways, this is the ``John Henry challenge''. Lattice methods,
despite a promising start, have been unable to resolve the problem
satisfactorily. In the analytic arena, the work of Buras, Bardeen and
Gerard$^{22}$ has taught us the right questions to ask and has stimulated a
modern way of approaching the problem, even if may feel that theirs is
not the final answer. 

The work on electromagnetic mass differences illustrates how this field
is developing. The early calculation of pion mass difference in the
chiral limit via dispersive sum rules is a benchmark, and when
combined with QCD indicates that high energy effects vanish in this
limit. In attempting to deal with on-shell pions and kaons, the
first approachs have been models of the intermediate energy region$^{23}$.
Subsequently more attention is being paid to a more systematic
treatment with more realistic matching with higher energies. My student
Antonio Perez$^{24}$ has done what I feel is the best job in this area by
using a dispersive treatment related to the Cottingham method. One can
identify all of the ingredients of the chiral limit approach, and
these come with identifiable on-shell corrections. The constraints of
QCD determine the high energy matching, those of chiral symmetry fix
the low energy structure, and data fix much of the intermediate
region. This provides a reasonably solid description of these matrix
elements.

Another recent calculation in this pioneering area is the B-parameter
work of Bijnens and Prades using the NJL model$^{25}$. 
While this is still a model
calculation, the approach is instructive. They use the model as a guide
to the matching in the intermediate energy region, which is the way
that one can remove the scale ambiguity of the calculations of chiral
loop corrections. The result has finite calculable corrections when
compared to the lowest order chiral prediction. While
the model is not a complete characterization of the intermediate
energy region, it does indicate what the physics is that we need to do
better if we are to calculate this amplitude reliably.   

I expect that this type of calculation will continue to develop
through an interplay of model and dispersive techniques. My
expectation is that we will rely increasingly on dispersive methods
for the final answers, perhaps with a bit of modeling thrown in to
account for the intermediate states that we cannot measure. These type
of calculations are also at the frontier of lattice work.

\section{Summary}

We certainly cannot claim to have completely tamed the strong
interactions yet. However we have made progress, in that both the high
energy and low energy regions have reliable methods that are now well
developed. In the intermediate energy region, we know the basic
physics and have ways to incorporate some of it into calculations. The
next step is to gain more control over these effects. My own vision of
how this can take place is to learn how to do a good job of modeling
the ingredients to dispersive calculations. Overall, I am optimistic
that we are still progressing in our calculational ability, and  I
personally find this progress to be one of the most interesting
aspects of the field at the moment. \\

\vspace{.2in}

\noindent {\large {\bf {References}}}.\\

\noindent 1) G. Kilcup\noindent , these proceedings.\\
2) It is just a curiosity that the papers that I chose for the
examples were all from 1967. Another pattern of selection would have
been to use all papers of Steven Weinberg who, in addition to another
well known 1967 paper and his effective field theory influence (see
Ref 9 below), produced seminal works in all the areas that I am
citing. \\
3) C. Bouchiat and Ph. Meyer, Phys. Lett. {\bf 25B}, 282 (1967).\\
4)For more references and a pedagogical background see 
 J. F. Donoghue, E. Golowich and B. R. Holstein, {\it Dynamics of the
Standard Model}, (Cambridge University Press, Cambridge, 1992).\\
5) J. Cronin, Phys. Rev. {\bf 161}, 1482 (1967).\\
6) S. Weinberg, Phys. Rev. Lett. {\bf 17}, 616 (1966).\\
T. Das, V. Mathur, and S. Okubo, Phys. Rev. Lett. {\bf 19}, 859
(1967).\\
7) T. Das, G. S. Guralnik, V. S. Mathur, F. E. Low, and J. E. Young,
Phys. Rev. Lett. {\bf 18}, 759 (1967).\\
8) J. F. Donoghue and E. Golowich, Phys. Rev. {\bf D49}, 1513 (1994).\\
9) Introductions to the ideas of effective field theory can
be found in:\\
S. Weinberg, {\em The Quantum Theory of Fields} 
(Cambridge University Press, Cambridge, 1995).\\
J.F. Donoghue, E. Golowich and B.R. Holstein, Ref 4 ,\\
H. Georgi, {\em Weak Interactions and Modern Particle Theory} 
(Benjamin/Cummings, Menlo Park, 1984). \\
S. Weinberg, Physica (Amsterdam) {\bf 96A}, 327 (1979). \\
J. Gasser and H. Leutwyler, Nucl. Phys. {\bf B250}, 465 (1985). \\
J.F. Donoghue, in {\em Effective Field Theories of the Standard Model}, 
ed. by U.-G. Meissner (World Scientific, Singapore, 1992) p. 3. \\
A. Cohen, Proceedings of the 1993 TASI Summer School, ed. by S. Raby 
(World Scientific, Singapore, 1994), p.53 . \\
Plenary talks by H. Leutwyler and S. Weinberg in {\em Proceedings of the 
XXVI International Conference on High Energy Physics}, Dallas 1992, 
ed. by J. Sanford (AIP, NY, 1993) pp. 185, 346.\\
D.B. Kaplan, Effective field theories, lectures at the 7th Summer
School in Nuclear Physics Symmetries, Seattle 1995, nucl-th/9506035.\\
A.V. Manohar, Effective field theories, Schladming lectures 1996,
hep-ph/9606222.  \\
10) Most of the references in Ref.[9] discuss chiral
perturbation theory. Other sources include:\\
J. F. Donoghue, Chiral symmetry as an experimental science,. in {\em
Medium Energy Antiprotons and the Quark-Gluon Structure of Hadrons},
ed. by R. Landua, J.-M. Richard and L. Klapish (Plenum, N.Y., 1991) p.
39 . \\
J. Bijnens, G. Ecker and J. Gasser, hep-ph/9411232.\\
A. Pich, Rept. Prog. Phys. {\bf 58}, 563 (1995).\\
G. Ecker, Erice lectures, hep-ph/9511412 . \\
11) S. Bellucci, J. Gasser and M. Sainio, Nucl. Phys. {\bf B423},80 (1994).\\
12) J. Bijnens, G. Colangelo, G. Ecker, J. Gasser and M. Sainio, Phys.
Lett. {\bf B374}, 210 (1996).\\
M. Knecht, B. Moussallam, J. Stern and N. Fuchs, Nucl. Phys. {\bf
B457}, 513 (1995).\\
J. Bijnens, G. Colangelo and J. Gasser Nucl. Phys, {\bf B427}, 427
(1994).\\
13) E. Golowich and J. Kambor, Nucl. Phys. {\bf B447}, 373 (1995),\\
K. Maltman, Phys. Rev. {\bf D53}, 2573 (1996).\\
J. Gasser and U. G. Meissner, Nucl. Phys. {\bf B357}, 90 (1991).\\
14) J. F. Donoghue, talks at the International school on effective
field theory Almunecar Spain 1995, and the International conference on
Hadrons and Nuclei Seoul 1995 hep-ph/9607351 and hep-ph/9506205. \\
15) D. Morgan and M. R. Pennington, Phys. Lett. {\bf B272}, 134 (1991).\\ 
J. F. Donoghue and B. R. Holstein, Phys. Rev. {\bf D48}, 137 (1993).\\
16) J. Bijnens and F. Cornet, Nucl. Phys. {\bf B296}, 557 (1988).\\
J. F. Donoghue, B. R. Holstein and Y. C. Lin, Phys. Rev. {\bf D37}, 2423 
(1988).\\
17) J. F. Donoghue and E. Golowich, Phys. Lett. {\bf 315}, 406 (1993).\\
18) J. F. Donoghue, C. Ramirez and 
G. Valencia, Phys.Rev.{\bf D39},1947,1989. \\
G. Ecker, J. Gasser, A. Pich, E. de Rafael, Nucl.Phys. {\bf B321},
311,1989. \\
19) J. Balog, Phys. Lett. {\bf B149}, 197 (1984). \\
A. Andrianov , Phys. Lett. {\bf B157}, 425 (1985).\\
P. Simic, Phys. Rev. {\bf D34}, 1903 (1986).\\
20) References can be found in J. Bijnens, Phys. Rept. {\bf 265}, 369
(1996). \\
21) J. Bijnens, Ch. Bruno and E. de Rafael, Nucl. Phys. {\bf B390},
501 (1993), \\
J. Bijnens, E. de Rafael, H-q. Zheng, Z. Phys. {\bf C62}, 437
{1994},\\
J. Bijnens and J. Prades, hep-ph/9409231\\
22) W. A. Bardeen, A. J. Buras, and J. M. Gerard, Phys. Lett. {\bf
B30}, 347 (1988); Nucl. Phys. {\bf B293}, 787 (1987). \\
23) J. Bijnens and E de Rafael, Phys. Lett. {\bf B273}, 483, (1991). \\
J. F. Donoghue, B. R. Holstein and D. Wyler, Phys. Rev. {\bf D47},
2089 (1993).\\
J. Bijnens, Phys. Lett. {\bf B306}, 343 (1993) .\\
24) A. Perez and J. F. Donoghue, in preparation. \\
25) J. Bijnens and J. Prades, Nucl. Phys. {\bf B444}, 523 (1995). 

\end{document}